# Chiral phonon magnetism from localized Raman excitation

**Authors:** Yuhan Wang[1] and Chuanshan Tian[1,2,*]

**Affiliations:**

[1]Department of Physics, State Key Laboratory of Surface Physics and Key Laboratory of Micro- and Nano-Photonic Structure (MOE), Fudan University, Shanghai 200433, China

[2]Institute of Quantum Science and Technology, Yanbian University, Yanji, Jilin 133002, China

* To whom correspondence should be addressed: cstian@fudan.edu.cn

## Abstract

Optically driven chiral phonons can induce transient magnetism in non-magnetic crystals, yet the microscopic relationship between phonon angular momentum and the resulting magnetic response remains unresolved. Here we distinguish the roles of pseudo-angular momentum and mechanical angular momentum in tetragonal α-$TeO_2$. Combining phase-resolved terahertz spectroscopy with magneto-optical Kerr measurements (MOKE), we reconstruct the motion of a doubly degenerate phonon and continuously control its angular momentum. The MOKE response follows the sign and magnitude of phonon angular momentum as well as its dynamics, whereas non-rotating phonons, including those with finite pseudo-angular momentum, produce no detectable magnetic signal. The MOKE rotation corresponds to an effective magnetic field of approximately 14 mT, nearly seven orders of magnitude above the estimate for spatially uniform ionic currents. Moreover, the Kerr response scales linearly with the emitted terahertz field, rather than quadratically as predicted by the uniform-current model. A localized Raman-excitation model involving Te lone-pair states accounts for this scaling and yields a comparable field magnitude with the MOKE data when electronic enhancement is included. These findings identify mechanical lattice rotation as the dominant dynamical variable for the observed magnetic response and suggest that localized Raman coupling contributes to its unexpectedly large magnitude.

## Introduction

Chiral phonons carry angular momentum[1-3] and have been implicated in a broad range of phenomena, including the phonon thermal Hall effect, the Einstein-de Haas effect, and ultrafast magnetization switching[4-17]. In the dynamical-multiferroic picture, circular ionic motion generates an effective magnetic field[11-13], thereby breaking time-reversal symmetry[17]. Recent experiments in $CeF_3$[12] and $SrTiO_3$[13] have observed substantial transient magneto-optical responses following excitation of circularly polarized phonons. Yet the effective magnetic fields inferred from these measurements can exceed estimates based on circulating ionic charges by orders of magnitude[13-15], suggesting a potentially important contribution from angular-momentum transfer to electronic degrees of freedom[13]. The microscopic relationship between phonon angular momentum and the resulting magnetization, however, remains unresolved[18].

A central challenge is that optical selection rules constrain phonon symmetry but do not uniquely determine the mechanical angular momentum (AM) of the excited lattice[2,3,19-21]. In crystals with discrete rotational symmetry, light-phonon interactions are governed by selection rules involving phonon pseudo-angular momentum (PAM) [2], a quantum number defined by the transformation of a phonon eigenstate under a discrete rotation[22,23]. Phonon AM, by contrast, arises from the real-space rotational motion of the ions and can vanish even for phonons with finite PAM[24,25]. Previous experiments have observed helicity-dependent magneto-optical responses following circular phonon excitation and interpreted them in terms of phonon angular momentum coupled to electronic degrees of freedom[13]. However, the symmetry of the optically selected phonons and their mechanical rotation were not independently disentangled, leaving unresolved which quantity governs the magnetic response[18]. Resolving this ambiguity is essential for understanding how angular momentum is transferred from light to phonons and subsequently to electrons, as well as the microscopic mechanism responsible for chiral-phonon-enhanced magnetism. This requires directly comparing phonon modes with distinct PAM-AM combinations while independently characterizing their lattice motion and magnetic response within the same crystal.

Tetragonal $\alpha$-$TeO_2$ offers such a comparison[26]. Its nondegenerate *A* and *B* phonons carry PAM indices of 0 and 2, respectively, but zero AM, whereas the doubly degenerate *E* phonon supports circularly polarized states with PAM = ±1 and opposite AM. The *B* mode thus provides an internal reference with finite PAM but no real-space ionic rotation, allowing its magnetic response to be compared directly with that of the rotating *E* mode. We combine symmetry-selective intrapulse difference-frequency excitation with phase-resolved vector electro-optic sampling[27,28] to reconstruct the *E*-mode trajectory, and use terahertz-induced Kerr[1] and magneto-optical Kerr measurements[13] to characterize the other phonon modes and the transient magnetic response, respectively.

Here, we show that the transient magneto-optical response follows both the magnitude and sign of the reconstructed AM as the *E*-mode trajectory is continuously tuned between opposite senses of rotation. By contrast, the long-lived *B* phonon produces no comparable magnetic response despite its finite PAM. This establishes real-space lattice rotation, rather than PAM alone, as the dynamical variable governing the observed magnetic response. The effective magnetic field inferred from the Kerr rotation reaches approximately 14 mT, nearly seven orders of magnitude larger than the approximately 2 nT predicted for spatially uniform ionic rotation. Moreover, at fixed phonon ellipticity, the Kerr response scales linearly with the emitted terahertz electric field as the pump power increases, rather than quadratically as expected from the uniform model[13-15]. We account for these observations using a localized Raman-excitation model involving polarizable Te lone-pair states[29-32], in which quantitative agreement requires an effective excitation volume of order one primitive cell. These findings identify mechanical lattice rotation as the dynamical origin of the observed magneto-optical response and suggest that localized electron-phonon coupling underlies its unexpectedly large magnitude.

## Main

Paratellurite $\alpha$-$TeO_2$ is a non-magnetic tetragonal crystal with point group $D_4$(422) and enantiomorphic space groups $P4_32_12$ and $P4_12_12$.[26,33] Its distorted Te-O units form a helical network along the crystallographic $c$ axis (Fig. 1a). Its 12-atom primitive cell gives rise to the zone-center optical vibrational representation (Fig. 1b and Extended Data Fig. 1)

$$\Gamma_{\text{opt, TeO2}} = 4A_1 + 4A_2 + 5B_1 + 4B_2 + 8E \qquad (1)$$

The $A_1$, $B_1$ and $B_2$ modes are Raman-active, whereas the $A_2$ mode is infrared-active, and the doubly degenerate $E$ mode is both Raman- and infrared-active. Under a fourfold rotation about $c$, these representations are characterized by PAM index $m = 0$ for $A$ mode, $m = 2$ for $B$ mode, and $m = \pm 1$ for the two circular combinations of the $E$ mode, with $m$ defined modulo four (Fig. 1b).

For the one-dimensional $A$ and $B$ representations, the atomic displacement is described by a single real eigenvector, so that the ions oscillate along a linear trajectory and carry no AM. By contrast, the two orthogonal coordinates of the doubly degenerate $E$ representation can form phase-shifted superpositions corresponding to circular or elliptical ionic trajectories in the $ab$ plane, thereby generating a finite AM along $c$ (Fig. 1b). The three symmetry classes thus realize distinct combinations of PAM and AM, $A$: (0,0), $B$: (2,0), $E$: (±1, ± L), with the $B$ mode providing an internal reference that separates finite PAM from mechanical rotation. The resonances near 1.8, 3.6 and 4.5 THz are assigned to *B*-, *E*- and *A*-symmetry modes, respectively (Fig. 1c).

We excite these phonons with a polarization-shaped optical pulse of approximately 50 fs centered at 1,030 nm. The high- and low-frequency spectral components of the pump are prepared with opposite helicities (Fig. 2a,b and Extended Data Fig. 2), allowing intrapulse Raman excitation of rotational-symmetry-selected phonons (Fig. 2a,b). For fourfold rotational symmetry, the corresponding selection rule is[2]

$$m = \sigma_1 - \sigma_2 - 4q \ (q = 0, \pm 1) \quad (2)$$

where $\sigma_{1,2}$ are the angular-momentum indices of the two participating optical-field components and $q$ is an integer. Transverse circular components have $\sigma = \pm 1$, while the component longitudinal to $c$ has $\sigma = 0$. Oblique incidence supplies both transverse and longitudinal fields, allowing $m = 0, \pm 1$, and 2 excitations, corresponding to the *A*, *E*, and *B* phonons in $\alpha$-$TeO_2$. Reversing the pump spectral helicities exchanges the $m = \pm 1$ of the *E* doublet, whereas the *B*-mode index remains $m = 2$ because $-2 \equiv 2$ (mod 4). Thus, optical selection rules determine the accessible phonon PAM, but not AM of the driven phonons.

We measure the two orthogonal components of the coherent *E* mode motion using phase-resolved vector electro-optic sampling. From the two measured polarization components of the emitted THz fields, we reconstruct the orthogonal phonon coordinates $Q_x(t)$ and $Q_y(t)$, and hence the time-dependent trajectory of the *E* phonon (Methods and Supplementary Information). Figures 2a and 2b present the reconstructed trajectory of elliptical THz fields in time domain of the *E* phonon, which winds in one direction in the *ab* plane. Reversing the pump helicities switches the excited *E* phonon from the $m = -1$ to $m = +1$ PAM channel. The emitted THz field consequently reverses its winding direction, demonstrating that the two circular *E* modes carry opposite AM. The emitted-field amplitude decays on a sub-picosecond timescale, with a characteristic time of approximately 0.87 ps.

The observed real-space rotation is not restricted to two discrete helicity states. As depicted in Fig. 2c, by continuously varying the spectral polarization of the excitation, the reconstructed trajectory evolves continuously from clockwise circular rotation, through increasingly elongated elliptical motion, and finally to counter-clockwise circular rotation. The axial AM of the *E* mode derived from the reconstructed trajectory, $L_z(t) \propto Q_x\dot{Q}_y - Q_y\dot{Q}_x$, therefore changes continuously in magnitude and sign (Extended Data Fig. 3a). The lattice rotation can therefore be obtained from the measured phonon response rather than inferred solely from the incident light helicity.

The non-rotating phonons provide a complementary dynamical reference to verify whether the real-space rotation is uniquely responsible for the magnetism in $\alpha$-$TeO_2$. We therefore characterized the *A* and *B* modes, which are Raman active and IR inactive, by time-resolved THz-induced Kerr effect (TKE) measurements[1] after intrapulse Raman excitation. The phonon trajectories for the two polarization directions are presented in Extended Data Fig. 4. The $A_1$ mode appears mainly under P-polarized detection, whereas the *B* mode appears mainly under S-polarized detection (weak orthogonal signals arising from crystal birefringence), confirming that both modes exhibit linear oscillations with zero phonon AM. We find that the temporal responses of these two modes are distinct from that of the *E* mode shown in Fig. 2. As shown in Fig. 3a, the $A_1$ mode at a 4.5 THz and *B* mode at 1.8 THz decay to 1/*e* of their maximum value by about 3.5 ps and tens of picoseconds, respectively.

We then compare these lattice dynamics with the time-resolved magneto-optical Kerr response (MOKE) measured under intrapulse Raman excitation as described above. Figure 3a presents the transient phonon-induced MOKE signal alongside the time-domain oscillations of the *A*, *B* and *E* modes. The Kerr response persists for approximately 1 ps, well beyond the 50-fs driving pulse, suggesting that the magnetic effect cannot be attributed to an instantaneous optical response, such as the inverse Faraday effect (IFE). (More discussion on IFE is given in Method and Extended Data Fig. 3d) Given that the phonon-induced magnetic field is proportional to phonon populations $|Q|^2$,[13] we compared the MOKE dynamics with $|Q|^2$ of the *A*, *B* and *E* modes in the inset in Fig. 3a. It reveals a clear temporal correspondence between the magnetic response and the *E* mode population, whereas the *A* and *B* modes exhibit much longer temporal dynamics. In particular, the *B* mode remains coherent after the transient MOKE response has disappeared despite its finite PAM. These observations establish that the magneto-optical response originates from the short-lived rotating *E* mode and not from the non-rotating *A* or *B* modes.

To further confirm the relation between phonon AM of the *E* mode and the magnetic response, we continuously varied the excitation polarization to tune the phonon $L_z$, following the approach shown in Fig. 2c. The MOKE signal extracted from Extended

Data Fig. 3a,b scales linearly with $L_z$ in both sign and amplitude (Fig. 3b). By contrast, it does not follow the same correspondence with the PAM of the phonon (Open scatters in Fig. 3b). In particular, a linear $E$ mode state ($L_z = 0$) produce no measurable magnetic signal within our experimental sensitivity. Thus, in $\alpha$-$TeO_2$ under these excitation conditions, the observed MOKE response must be governed by the real-space rotational motion (AM) of the lattice, rather than by PAM alone. Magnetic response requires the breaking of time-reversal symmetry. A lattice state with nonzero AM breaks time-reversal symmetry, whereas finite PAM does not necessarily do so. (see details in Methods).

The magnitude of the response places an additional constraint on the microscopic origin of phonon-induced magnetic response. From the measured Kerr rotation, we calculate the effective magnetic field $\boldsymbol{B}$ and magnetic moment $\boldsymbol{m}$ using the Verdet constant $V$ of $\alpha$-$TeO_2$, [34,35] together with the effective probe interaction length $l$[13]:

$$\boldsymbol{B} = \frac{\theta_\mathrm{K}}{Vl}, \ \boldsymbol{m} = \frac{\theta_\mathrm{K} V_c}{\mu_0 V l} \tag{3}$$

Here, $\theta_\mathrm{K}$ is the Kerr rotation angle measured by MOKE, $V_c$ is the unit cell volume. $\boldsymbol{B}$ is found to be approximately 14 mT, corresponding to a magnetic moment $\boldsymbol{m} \approx 0.2\ \mu_\mathrm{B}$. (see details in SI) Alternatively, we can also estimate $\boldsymbol{B}$ from the orthogonal normal coordinates $Q_x$ and $Q_y$ deduced from the radiated THz field $\boldsymbol{E}$. Assuming these quantities are uniformly distributed in bulk, $\boldsymbol{E}$, $\boldsymbol{B}$ and $\boldsymbol{Q}$ follow the relation (see Method)

$$\boldsymbol{E_i} = \frac{\sqrt{\mu_0/\varepsilon_0}}{n_0^2 V_c} \Omega q_\mathrm{eff} l_\mathrm{eff} Q_i, \ \boldsymbol{B} = s g_B \frac{\Omega q_\mathrm{eff} \mu_0}{V_c} Q_0^2 = s g_B \frac{\varepsilon_0 n_0^4 V_c}{\Omega q_\mathrm{eff} l_\mathrm{eff}^2} \boldsymbol{E}^2 \tag{4}$$

where $Q_i$ ($i = x,y$) denotes the relevant normal-coordinate displacement; $q_{\mathrm{eff}}$ and $l_{\mathrm{eff}}$ are the effective charge and interaction length, respectively, and $n_0$ is the refractive index in the THz frequency range. $s$ is the phonon helicity, $g_B$ is a geometrical factor, and $\Omega$ is the phonon frequency. Inserting the known parameters and the measured THz field $\boldsymbol{E}$ in Eq. (4) gives an ionic-current generated $\boldsymbol{B}$ of approximately 2 nT (see SI). This value is nearly seven orders of magnitude below the Kerr-equivalent estimate. Even when angular momentum transfer from phonons to electrons is taken into account[13-15], which raising the estimate to approximately 10 μT, the resulting field remains three orders of magnitude smaller than the experimental value.

We notice that the spatially uniform model in Eq. (4) predicts a quadratic dependence of $\boldsymbol{B}$ on $\boldsymbol{E}$. However, our measurements in $\alpha$-$TeO_2$ reveal a linear dependence (Fig. 3c), contradictory to the prediction assuming uniform ionic current. In $\alpha$-$TeO_2$, the stereochemically active Te ($5s^2$) lone-pair states have localized electronic character[29-32] and have been shown to contribute substantially to the off-resonant Raman response [32,33,36]. This motivate a local Raman-excitation picture in which lattice motion is initially excited within atomic-scale regions associated with these lone-pair states (Fig. 3d). Taking the characteristic $E$-mode response time to be approximately 1 ps and the phonon group velocity calculated from the chiral phonon dispersion curve[26] to be $v_g \sim 0.8\ \mathrm{km \cdot s^{-1}}$, we obtain a propagation length of approximately 0.8 nm. This sub-nanometer scale suggests that an initially localized lattice excitation and its associated magnetic response could remain spatially confined over the observed transient.

Within this picture, each local excitation[37,38] event is assumed to deposit an energy $\Delta E = \hbar\Omega$ into the lattice, setting a characteristic initial displacement $Q_0$, while the excitation density $n_{\mathrm{ph}}$ determines the macroscopic phonon response. For the $E$ mode at $\Omega = 2\pi \times 3.66$ THz, approximating the effective vibrating mass by the oxygen mass gives $Q_0 = \sqrt{\frac{2\Delta E}{M\Omega^2}} \approx 19$ pm. The macroscopic lattice response then arises from the spatial average of these localized excitation excitations (see Method for details):

$$\boldsymbol{E} = \frac{\sqrt{\mu_0/\varepsilon_0}}{n_0^2}\Omega q_{\mathrm{eff}} l_{\mathrm{eff}} Q_0 n_{\mathrm{ph}},\ \boldsymbol{B} = s g_B \Omega q_{\mathrm{eff}} \mu_0 Q_0^2 n_{\mathrm{ph}} = s g_B \frac{n_0^2}{c l_{\mathrm{eff}}} Q_0 \boldsymbol{E} \tag{5}$$

Including the assumed angular-momentum transfer to electronic degrees of freedom[13] or non-Maxwellian enhancement[39], the calculated $\boldsymbol{B}$ reaches approximately 10 mT, comparable to the value of 14 mT from MOKE measurement. Crucially, Eq. (5) predicts a linear dependence of the magnetic field on the THz electric field $\boldsymbol{B} \propto \boldsymbol{E}$, rather than its square $\boldsymbol{B} \propto \boldsymbol{E}^2$, consistent with the measured scaling (Fig.3c).

## Conclusion and outlook

Our measurements distinguish the phonon symmetry selected by light from the mechanical angular momentum associated with transient magnetism in $\alpha$-$TeO_2$. The MOKE response follows the reconstructed rotation of the $E$ mode, whereas the $B$ mode produces no comparable signal despite carrying finite PAM. The inferred field magnitude and its linear scaling with emitted terahertz amplitude cannot be explained by a spatially uniform ionic-current model. A localized Raman-coupling model involving Te lone-pair states accounts for both observations, although the excitation volume and electronic angular-momentum-transfer pathway remain to be established independently. Resolving the mode-specific Raman coupling and the orbital and spin contributions to the magneto-optical response will provide a more stringent test of this microscopic picture.

More broadly, our Raman excitation scheme enables efficient angular-momentum transfer to the lattice by tuning the frequency difference between two optical fields into resonance with a phonon mode. In crystals with $C_n$ ($n$ = 3,4,6) rotational symmetry, as well as in quasicrystals with n $\geq$ 3, $E$ mode phonons can be efficiently excited through phonon-resonant Raman scattering and carry PAM = $\pm$1. The resulting AM may provide a route to optically controlling transient magnetic responses through coupling between phonons and electrons. In systems with even-fold rotational symmetry, $C_n$ with even $n$ supports $B$-mode phonons with PAM = $n$/2. Although these modes carry no AM, their excitation transfers optical angular momentum to the whole crystal, instead of an individual phonon mode. By contrast, $A$ mode phonons in crystals with all allowed rotational symmetries ($n$ = 1,2,3,4,6), or in quasicrystals, carry neither PAM nor AM and therefore do not absorb angular momentum from the optical field. Therefore, these symmetry-dependent excitation pathways provide a framework for designing light-driven phonon control of magnetic responses, with potential applications in the ultrafast, mode-selective manipulation of magnetic states.

## Main references

# Methods

## Terahertz difference-frequency spectroscopy

Terahertz difference-frequency spectroscopy (THz-DFS) was performed using a Yb:KGW laser (PHAROS, Light Conversion) operating at a central wavelength of 1030 nm and delivering 190 fs pulses at a repetition rate of 100 kHz. The pulses were externally compressed to 26 fs using a home-built compressor[40]. After broadening due to group-velocity dispersion (GVD) in the 1-mm-thick quartz waveplate and subsequent polarization optics, the 26-fs pulse reaches the sample with a duration of approximately 50 fs. The output beam was split by a 9:1 beam splitter, with 90% of the pulse energy directed to the sample to generate terahertz radiation via second-order difference-frequency generation, and the remaining 10% used as a probe for detection.

The emitted terahertz electric field was detected in the time domain by electro-optic sampling, enabling phase-resolved detection of both amplitude and polarization of the field. This allows direct reconstruction of the vectorial terahertz waveform associated with the lattice response. The measured difference-frequency spectra were normalized using GaP as a reference sample. In the frequency range below 5 THz, GaP exhibits an essentially dispersionless second-order nonlinear susceptibility, ensuring reliable normalization (see SI and Ref.27 and Ref.28 for further details on the experimental setup and data processing).

## Circularly Polarized Terahertz Difference Frequency Spectroscopy

Circularly polarized terahertz difference-frequency spectroscopy was implemented using intrapulse difference-frequency generation from a broadband femtosecond laser pulse centered at 1030 nm, with a spectral bandwidth of approximately 80 nm. A 1-mm-thick quartz plate was used as a dispersive waveplate, providing a wavelength-dependent retardance varying continuously from 0.75 to 0.25 across the 1006-1064 nm spectral range (Extended Data Fig. 2). As a result, the high- and low-frequency components within the same pulse acquire opposite circular polarizations, enabling circularly polarized difference-frequency excitation within a single optical beam.

In this configuration, the helicity difference between the two frequency components determines the pseudo-angular momentum (PAM) transferred from the optical field to the lattice, thereby selecting phonon modes according to symmetry-imposed selection rules. Importantly, this transfer of PAM does not by itself guarantee real-space angular momentum (AM), which must be established independently through the emitted terahertz field.

The temporal walk-off between the two polarization components, arising from the birefringence between the ordinary and extraordinary axes of 1-mm-thick quartz, is less than 30 fs, and the duration of the laser pulse incident on the sample is approximately 50 fs, which is significantly shorter than the period of the emitted terahertz radiation (~300 fs), ensuring strong temporal overlap of the excitation pathways. Consequently, the phonons are coherently driven by the combined circularly polarized field, rather than by two independent linearly polarized excitations, which is essential for generating well-defined angular-momentum-carrying lattice motion.

## Non-magnetic THz-induced Kerr effect (Raman phonons) and Magneto-optical Kerr measurements

The non-magnetic THz-induced Kerr effect (TKE)[1] was measured in a transmission geometry, whereas the magneto-optical Kerr effect (MOKE) was measured in reflection. In the reflection geometry, the contribution from the non-magnetic THz-induced Kerr effect was negligible and below the detection limit. The 515-nm probe beam was generated by frequency doubling a split portion of the 1030-nm pump beam. Its spot size (~200 $\mu$m) was smaller than that of the pump beam (~500 $\mu$m).

We first performed time-resolved measurements of the non-magnetic THz-induced Kerr effect to characterize the dynamics of the Raman-active *A* and *B* modes. The excitation scheme was identical to that used for the circularly polarized THz difference-frequency spectroscopy described above. A quarter-wave plate and a Wollaston prism were used to detect the pump-induced change in the polarization state of the probe beam. By using S-polarized (in-plane) and P-polarized (out-of-plane) probe beams, we distinguished the contributions of the *B* mode at 1.8 THz and the *A* mode at 4.5 THz, respectively. We then performed MOKE measurements for eight different polarization combinations of the excitation pulses and recorded the transient polarization rotation of the probe beam.

## Ionic Current Model

We first estimate the magnetic field generated by a spatially uniform circulating ionic current associated with the *E* mode phonon. In this classical model, each unit cell is approximated as a rotating charge.

Let the displacement of atom $\kappa$ in unit cell $l$ be

$$\boldsymbol{u}_{l\kappa} = u_{\kappa x}\hat{x} + u_{\kappa y}\hat{y} \tag{6}$$

The ionic current density generated by the coherent phonon is

$$\boldsymbol{J} = \frac{1}{V_c}\sum_\kappa Z_\kappa^* e \dot{u}_\kappa \tag{7}$$

where $Z_\kappa^*$ is the mode-projected effective charge, $e$ is the elementary charge and $V_c$ is the primitive-cell volume. Projecting onto the two orthogonal $E$ mode coordinates gives

$$J_x = \frac{e}{V_c} Z_x \dot{Q}_x,\ J_y = \frac{e}{V_c} Z_y \dot{Q}_y \tag{8}$$

where $Z_x$ and $Z_y$ are the mode effective charges and $Q_x$, $Q_y$ are the two orthogonal normal coordinates. The corresponding AM is

$$L_z = \sum_\kappa M_\kappa \left(u_{\kappa x} \dot{u}_{\kappa y} - \dot{u}_{\kappa x} u_{\kappa y}\right) \tag{9}$$

For the $E$ mode motion, the angular momentum has the general form

$$L_z = s\Omega Q_0^2 M \tag{10}$$

where $s$ denotes the ellipticity parameter, $\Omega$ is the phonon frequency, and $M$ is the effective mass. The polarization associated with the coherent phonon is

$$P_i = \frac{eZ_i}{V_c} Q_i = \frac{q_{\text{eff}}}{V_c} Q_i \tag{11}$$

The emitted THz field is proportional to the phonon polarization. We write

$$\boldsymbol{E}_i = \frac{\sqrt{\mu_0/\varepsilon_0}}{n_0^2} \Omega l_{\text{eff}} P_i = \frac{\sqrt{\mu_0/\varepsilon_0}}{n_0^2 V_c} \Omega q_{\text{eff}} l_{\text{eff}} Q_i \tag{12}$$

where $n_0$ is the refractive index at the phonon frequency.

The magnetic field generated by the circulating ionic charge is determined by the curl of the current distribution. Let $\boldsymbol{m}$ be the magnetic moment associated with one primitive cell. The primitive-cell number density is therefore $n_c = 1/V_c$. The corresponding magnetization is $\boldsymbol{M} = n_c\boldsymbol{m}$. Within the effective-medium approximation and neglecting demagnetization effects, the magnetic field associated with this magnetization is

$$\boldsymbol{B} = \mu_0 n_c \boldsymbol{m} = \frac{\mu_0}{V_c}\boldsymbol{m} \tag{13}$$

For a circular or elliptical ionic trajectory, the magnetic moment can be estimated from $\boldsymbol{m} = IA$, where $I$ is the effective circulating current and $A$ is the area enclosed by the trajectory. For an effective charge $q_{\mathrm{eff}}$ moving with angular frequency $\Omega$ and displacement amplitude $Q_0$, the magnetic moment is

$$\boldsymbol{m} = s g_B \Omega q_{\mathrm{eff}} Q_0^2 \tag{14}$$

Substituting Eq. 14 into Eqs. 12 and 13, we obtain

$$\boldsymbol{B} = s g_B \frac{\Omega q_{\mathrm{eff}} \mu_0}{V_c} Q_0^2, \; \boldsymbol{B} = s g_B \frac{\varepsilon_0 n_0^4 V_c}{\Omega q_{\mathrm{eff}} l_{\mathrm{eff}}^2} \boldsymbol{E}^2 \tag{15}$$

Equation 15 therefore gives the prediction of the spatially uniform ionic-current model. In this picture, the magnetic field amplitude scales quadratically with the emitted THz-field amplitude.

## Localized Raman excitation model

To account for the large discrepancy between the uniform ionic-current estimate and the Kerr-inferred effective field, we consider a localized Raman excitation model. In this picture, the Raman process is initially confined to a spatial region smaller than or comparable to the size of a unit cell, while the macroscopic coherent phonon response results from the collective contribution of many local excitation events within the illuminated volume. This picture is motivated by the localized Te-derived lone-pair states in $\alpha$-$TeO_2$, which may provide a localized contribution to the off-resonant Raman polarizability.

Let $N_{\mathrm{ph}}$ denote the number of local excitation events within an illuminated volume $V$. The corresponding excitation density is

$$n_{\mathrm{ph}} = \frac{N_{\mathrm{ph}}}{V} \tag{16}$$

If each excitation event deposits an energy $\Delta \mathrm{E} = \hbar\Omega$, the resulting local displacement $Q_0$ is determined by

$$Q_0 = \sqrt{\frac{2\hbar}{M\Omega}} \tag{17}$$

Here $M$ is the effective mass. The local polarization and magnetic moment then follow the same ionic-current relation

$$\boldsymbol{p} = q_{\mathrm{eff}} Q_0,\ \boldsymbol{m} = s g_B \Omega q_{\mathrm{eff}} Q_0^2 \tag{18}$$

The emitted THz electric field and the corresponding magnetic field are

$$\boldsymbol{E} = \frac{\sqrt{\mu_0/\varepsilon_0}}{n_0^2} \Omega q_{\mathrm{eff}} l_{\mathrm{eff}} Q_0 n_{\mathrm{ph}},\ \boldsymbol{B} = s g_B \Omega q_{\mathrm{eff}} \mu_0 Q_0^2 n_{\mathrm{ph}} \tag{19}$$

Eq. 19 gives

$$\boldsymbol{B} = s g_B \frac{n_0^2}{c l_{\mathrm{eff}}} Q_0 \boldsymbol{E} \tag{20}$$

Equation 20 is the prediction of the localized Raman excitation model. In contrast to the spatially uniform model, the magnetic field amplitude now scales linearly with the emitted THz-field amplitude. The radiated electric field remains unchanged, whereas the magnetic field is enhanced by a factor of $1/n_{ph}V_c$, corresponding to the inverse of the number of excitations per unit cell, because the angular momentum and magnetic moment are initially concentrated within localized excitation events rather than distributed uniformly throughout the unit-cell volume.

## Time-reversal origin of the magnetic response

The different roles of PAM and AM can be understood directly from time-reversal symmetry. For the lattice-displacement operator $\hat{u}_{l\kappa}$, where $l$ labels the unit cell and κ labels the atom, the real AM of a lattice mode is obtained from the mass-weighted ionic motion,

$$\hat{L}_{\mathrm{AM}} = \sum_{l\kappa} M_\kappa \, \hat{u}_{l\kappa} \times \dot{\hat{u}}_{l\kappa} \tag{21}$$

Time reversal leaves the displacement unchanged but reverses the velocity. Hence $T\hat{L}_{\mathrm{AM}}T^{-1} = -\hat{L}_{\mathrm{AM}}$, while magnetization and magnetic field are likewise odd under time reversal. A rectified magnetic response in the initially time-reversal-symmetric $\alpha$-$TeO_2$ crystal therefore requires a driven lattice state containing a time-reversal-odd degree of freedom.

For the doubly degenerate $E$ mode, the two circular states PAM = ±1 are exchanged by time reversal. Their population imbalance and the resulting Lz are therefore time-reversal odd. Reversing the excitation helicity exchanges these states, simultaneously reversing both $L_z$ and the observed magnetic response. PAM, on the other hand, transforms differently. Under time reversal, $m \rightarrow -m$ modulo the rotational order. The $E$ mode states are therefore distinct time-reversed partners, whereas the $B$ mode value $m = 2$ is self-conjugate. The PAM of the B mode therefore carries no time-reversal-odd sign, while its real eigenvector gives zero cycle-averaged AM. PAM thus specifies the rotational symmetry channel selected in the light-matter interaction, whereas AM characterizes the time-reversal-odd lattice dynamics associated with the observed rectified magnetic response. In $\alpha$-$TeO_2$, under the present time-reversal-symmetric conditions, it is this real-space rotation rather than PAM alone that provides the relevant dynamical descriptor of phonon-induced magnetism.

## Inverse Faraday effect

In our impulsive difference-frequency scheme, the high- and low-frequency components of the optical field carry equal and opposite angular momenta and therefore reduce the contribution expected from a conventional inverse Faraday effect (IFE). When the electronic transitions are far from resonance, the net angular momentum transferred to the electronic system vanishes. Furthermore, as shown in Extended Data Fig. 3d, the MOKE signal induced by a single circularly polarized optical field is weaker than that generated by the difference-frequency excitation involving distinct high- and low-frequency components. This further suggests that the conventional IFE makes only a limited contribution to the observed magnetic response.

## Samples and Experimental Environment

$\alpha$-$TeO_2$ single crystals (0.5mm thickness) were commercially purchased from HF-Kejing. The crystals were z-cut and double-side polished. All measurements were performed at room temperature under ambient conditions. The sample was periodically translated during acquisition when necessary to avoid cumulative damage or local heating. No external magnetic field was applied unless otherwise stated. To achieve a higher SNR, a 5-mm-thick crystal (Fig 2c,3b,3c and Extended Data Fig. 3) was used to quantitatively compare the MOKE and THz-DFS signals under different pump conditions. This thickness introduces some temporal broadening for THz-DFS, making direct time-domain comparisons unsuitable; however, the signal amplitudes remain unaffected.

### Data availability

The datasets generated and analyzed during the current study are available from the corresponding author on reasonable request.

### Code availability

All custom codes used for the data processing and numerical simulations are available from the corresponding author upon reasonable request.

## Method references

## Acknowledgements

We thank Y. Ron Shen for valuable discussions. This work was supported by the

National Natural Science Foundation of China (Grant Nos. 12125403, 12293053, 12221004, and 12250002), the Shanghai Municipal Science and Technology (Grant Nos. 23dz2260100 and 23JC1400400) and the National Key Research and Development Program of China (Grant Nos. 2021YFA1400503 and 2021YFA1400202).

## Author contributions

Y.-H.W. and C.-S.T. conceived the study. Y.-H.W. performed the terahertz difference-frequency spectroscopy, terahertz Kerr, and magneto-optical Kerr measurements. Y.-H.W. and C.-S.T. analyzed the data. Y.-H.W. and C.-S.T. drafted finalized the manuscript.

## Competing interest declaration

The authors declare no competing interests.

## Corresponding author

Correspondence to Chuanshan Tian

## Supplementary information

Supplementary Information.pdf

## Figures and captions

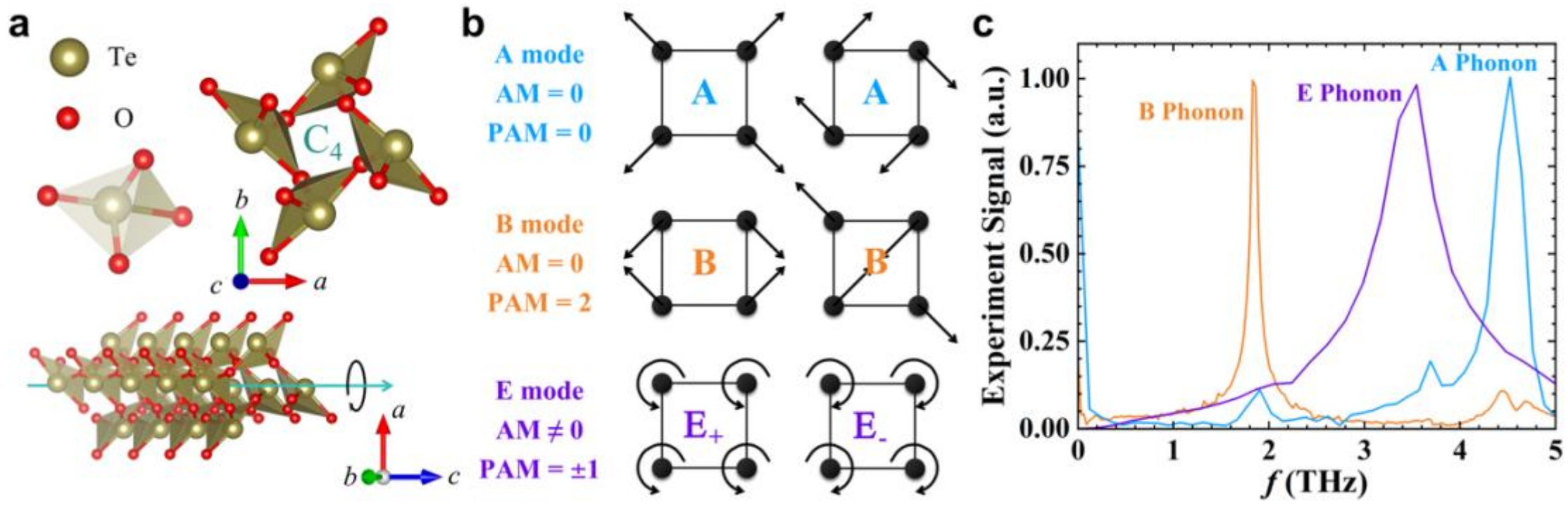

**Figure 1: Crystal structure and symmetry classification of phonon PAM and AM in *α*-TeO₂. a,** Crystal structure of tetragonal *α*-$TeO_2$, showing distorted Te-O units and their helical arrangement along the c axis (fourfold rotation axis). Gold and red spheres denote Te and O atoms. **b,** Displacement patterns of *A*, *B* and *E* modes under fourfold rotation. *A* and *B* modes have PAM = 0 and 2, respectively, but zero AM, whereas the two circular *E* mode combinations have PAM = ±1 and opposite AM. PAM is defined modulo 4. **c,** Normalized spectra showing the *B* (TKE), *E* (THz-DFS) and *A* (TKE) modes resonances at approximately 1.8, 3.6 and 4.5 THz, respectively.

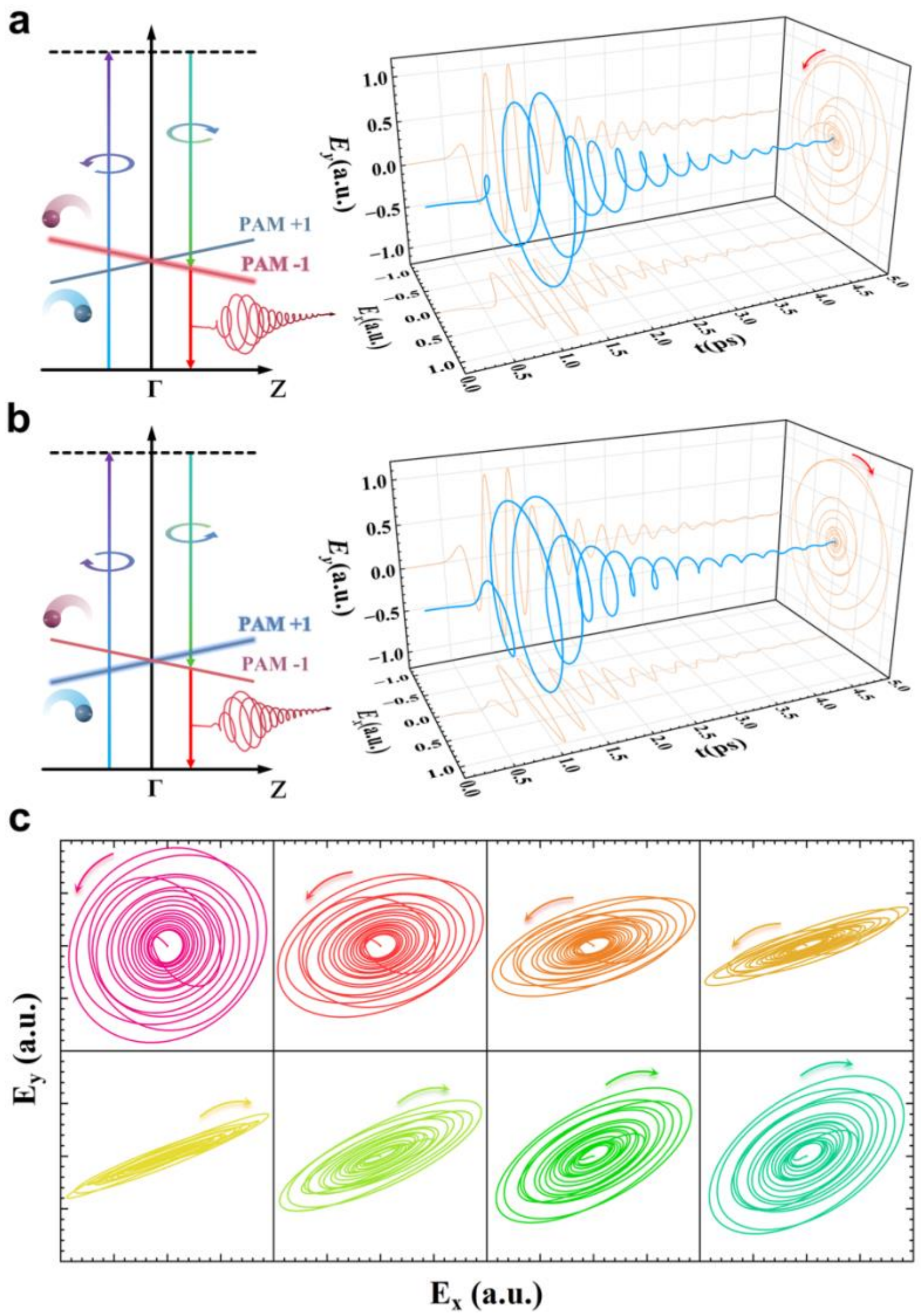

**Figure 2: Polarization-dependent excitation and vector characterization of the E-mode response. a, b,** Intrapulse difference-frequency excitation with opposite spectral-helicity configurations (left) and the corresponding phase-resolved vector THz waveforms (right). Reversing the spectral helicities switches the selected E-mode PAM channel between m = −1 and m = +1, accompanied by a reversal of the winding direction of the emitted terahertz field. **c,** Parametric trajectories of the orthogonal terahertz electric-field components, $\mathbf{E_x}$ and $\mathbf{E_y}$, for eight excitation-polarization settings. The trajectories illustrate continuous control of the field ellipticity and rotation sense, from nearly circular motion through increasingly elliptical trajectories

to rotation in the opposite direction. Arrows indicate the direction of temporal evolution.

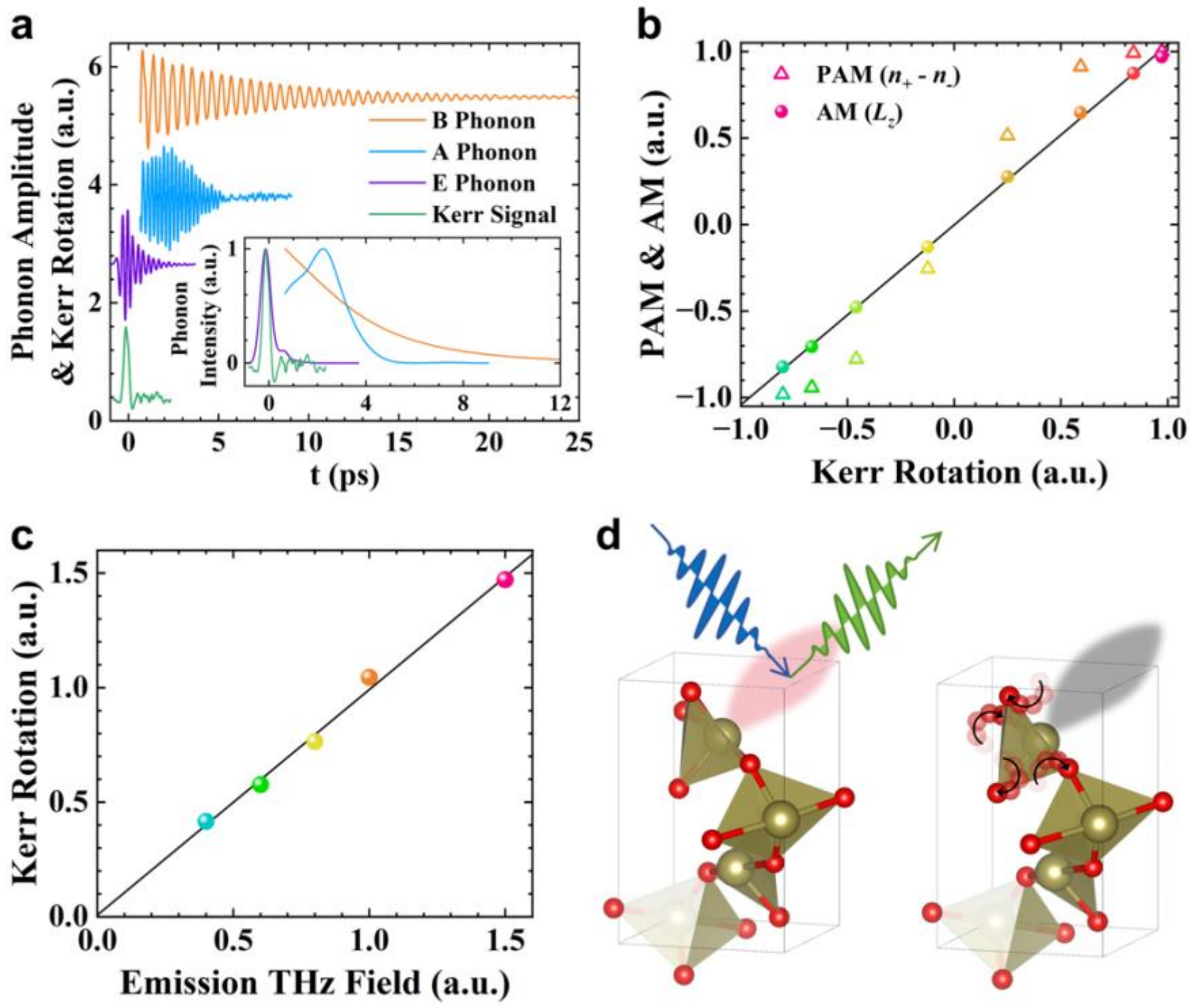

**Figure 3: AM-dependent magneto-optical response. a,** Time-domain responses of the *B* (orange), *A* (blue) and *E* (purple) phonons, together with the transient Kerr signal (green). The *B* and *A* modes response were measured by TKE spectroscopy, whereas the *E*-mode dynamics were obtained from the THz-DFS. The Kerr signal was measured by MOKE. Inset, Normalized phonon-response envelopes and the transient Kerr response, highlighting their distinct temporal dynamics. **b,** PAM population imbalance, $n_+ - n_-$ (triangles), and the trajectory-derived AM, $L_z$ (circles) plotted against MOKE signals for different excitation-polarization settings. **c,** Kerr rotation as a function of the emitted terahertz electric-field amplitude for different pump powers. **d,** Schematic illustration of the proposed localized Raman-excitation model. Intrapulse difference-frequency excitation drives localized Raman processes

associated with Te-derived lone-pair states, inducing local lattice displacements and ionic circulation.

# Extended data

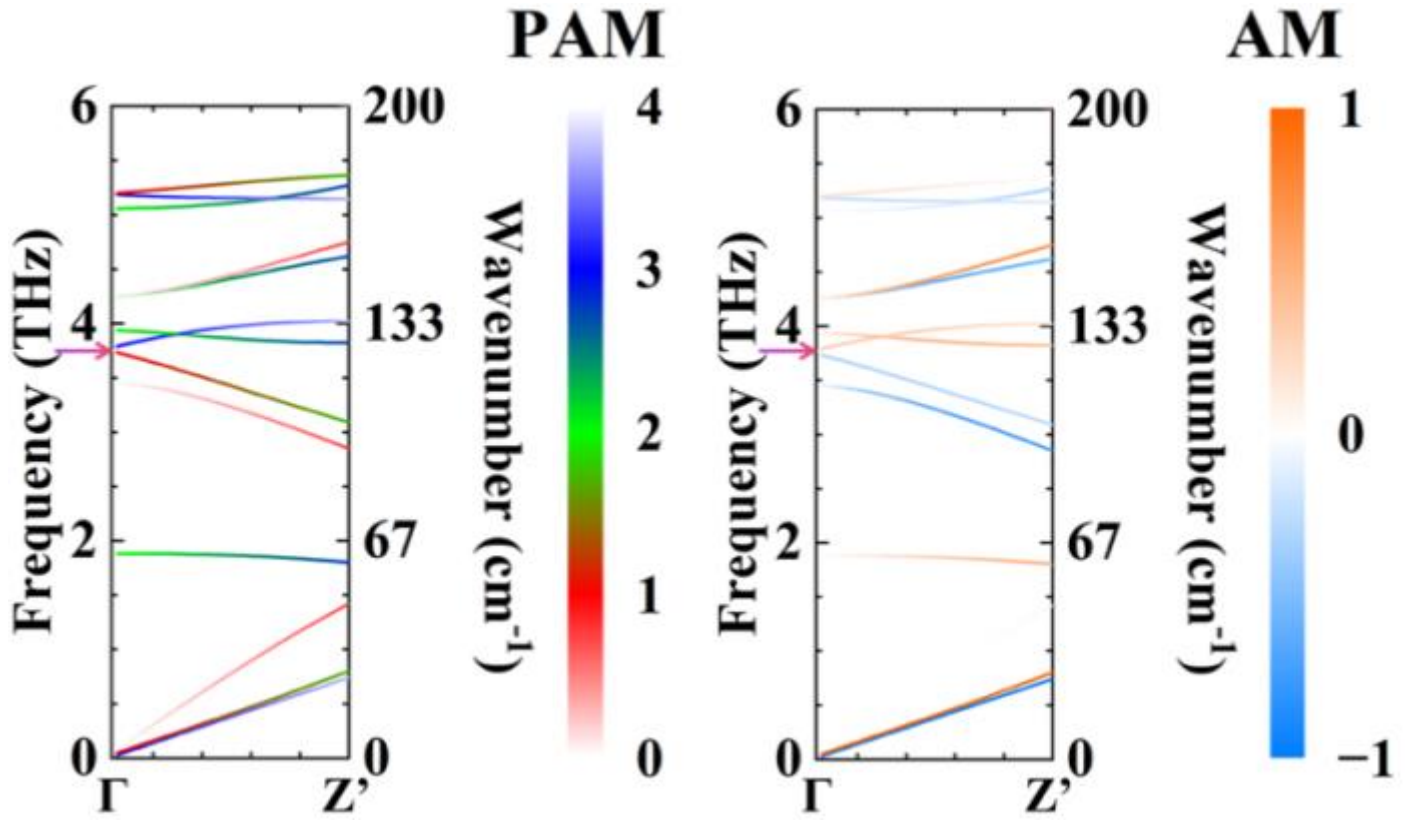

**Extended Data Fig. 1: Calculated phonon dispersion[26] with PAM and AM.** Calculated phonon dispersion along the Γ-Z′ direction, with Z′ corresponding to half the wavevector of Z. The left and right panels encode the PAM (mod 4) and AM, respectively. Magenta arrows highlight the E-mode phonon near 3.66 THz at the Γ point.

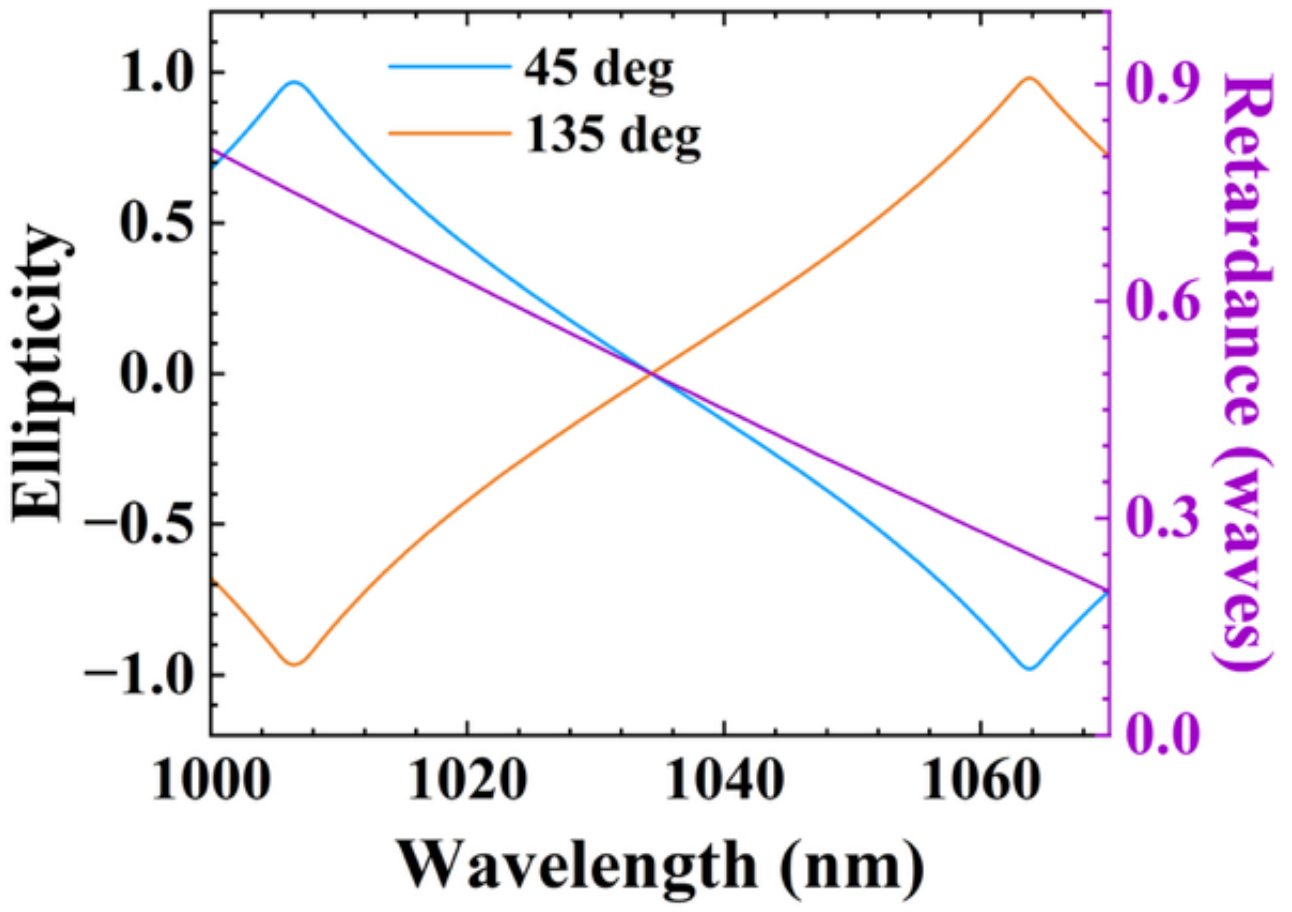

**Extended Data Fig. 2: Ellipticity of the pump (left) and retardance of the 1-mm quartz plate (right).** Polarization calibration using a 1-mm quartz plate (commercially purchased from Thorlabs, WPMQ10M-1064). Ellipticity of the transmitted optical field as a function of wavelength for two orientations of the fast axis of a 1-mm-thick z-cut quartz plate, set at 45° (blue line) and 135° (orange line) with respect to the x-axis. The purple curve represents retardance of 1-mm quartz plate as a function of wavelength.

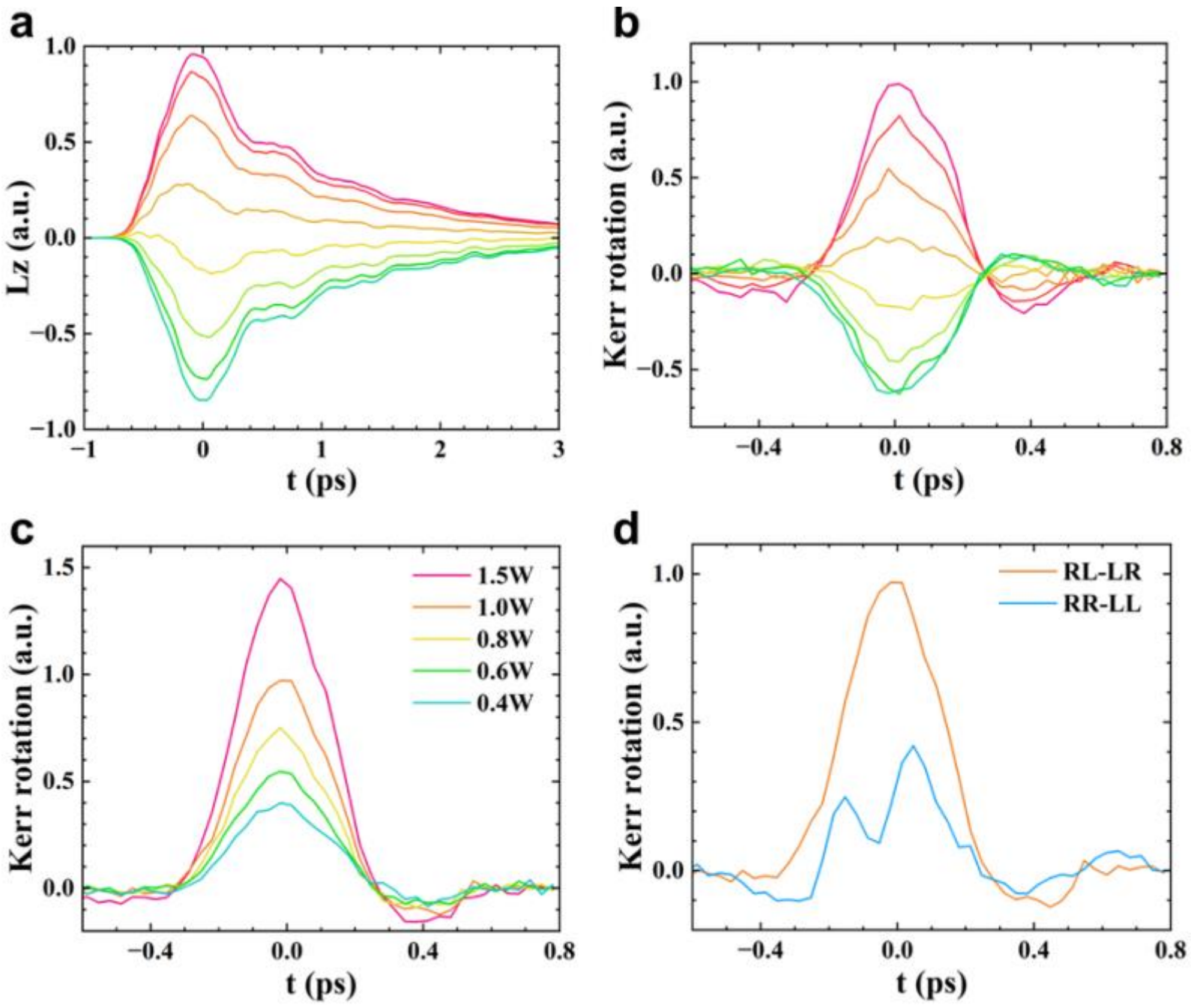

**Extended Data Fig. 3: Temporal evolution of the reconstructed E mode AM and the magneto-optical response. a,** Time-dependent out-of-plane mechanical angular momentum, $L_z(t) \propto Q_x\dot{Q}_y - Q_y\dot{Q}_x$, reconstructed for different excitation-polarization settings. **b,** Corresponding time-resolved Kerr rotation for the same excitation conditions. The curve colors in a and b correspond to the excitation-polarization settings represented in Fig. 2c and Fig. 3b. **c,** Time-resolved Kerr rotation measured at pump powers of 0.4, 0.6, 0.8, 1.0 and 1.5 W, as indicated. The curve colors in c correspond to Fig. 3c. **d,** Time-resolved Kerr rotation for excitation with opposite and identical angular momenta of the high- and low-frequency components. RL denotes the combination of a right-handed circularly polarized high-frequency component and a left-handed circularly polarized low-frequency component. The other labels follow the same convention. The signals shown are the differential signals between excitation configurations with opposite circular-polarization combinations. The weaker response under excitation with identical angular momenta (pure circular polarization) indicates that the inverse Faraday effect is not the dominant contribution.

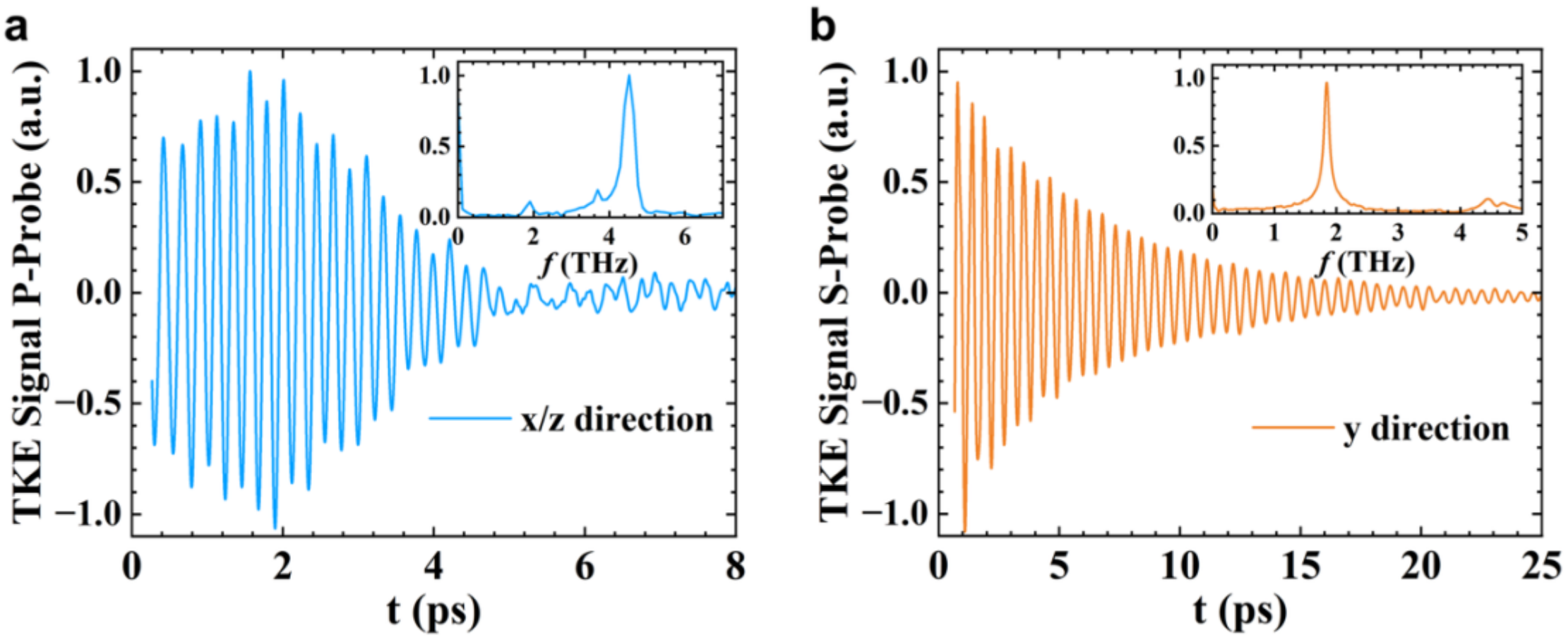

**Extended Data Fig. 4: Time-domain and frequency-domain terahertz Kerr effect (TKE) response. a,** Time-resolved and frequency-domain (inset) TKE response of the x/z direction vibrations with a P-polarized probe, which can be assigned to the A mode based on the vibrational frequency and polarization direction. **b,** Time and frequency resolved TKE response of the in-plane B mode measured with a S-polarized probe.